# Beyond steric selectivity of ions using angstrom-scale capillaries

Solleti Goutham[1,2], Ashok Keerthi [2,3], Abdulghani Ismail[1,2], Ankit Bhardwaj[1,2], Hossein Jalali,[4] Yi You[1,2], Yiheng Li,[5] Nasim Hassani,[4] Haoke Peng,[5] Marcos Vinicius Surmani Martins[1,2], Fengchao Wang,[5] Mehdi Neek-Amal,[4,6] Boya Radha[1,2,*]

[1]Department of Physics and Astronomy, School of Natural Sciences, The University of Manchester, Manchester M13 9PL, United Kingdom
[2]National Graphene Institute, The University of Manchester, Manchester M13 9PL, United Kingdom
[3]Department of Chemistry, School of Natural Sciences, The University of Manchester, Manchester M13 9PL, United Kingdom
[4]Department of Physics, Shahid Rajaee Teacher Training University, 16875-163 Lavizan, Tehran, Iran
[5]Chinese Academy of Sciences Key Laboratory of Mechanical Behavior and Design of Materials, Department of Modern Mechanics, University of Science and Technology of China, Hefei, 230027, China.
[6]Department of Physics, University of Antwerp, Groenenborgerlaan 171, Antwerp B-2020, Belgium

* Correspondence to be addressed to: radha.boya@manchester.ac.uk

**Abstract**
**Ion-selective channels play a key role in physiological processes and are used in many technologies. While biological channels can efficiently separate same-charge ions with similar hydration shells, it remains a challenge to mimic such exquisite selectivity using artificial solid-state channels. Although, there are several nanoporous membranes that show high selectivity with respect to certain ions, the underlying mechanisms are based on the hydrated ion size and/or charge. There is a need to rationalize the design of artificial channels to make them capable of selecting between similar-size same-charge ions, which in turn requires understanding of why and how such selectivity can occur. To address this issue, we study angstrom-scale artificial channels made by van der Waals assembly, which are comparable in size with typical ions and carry little residual charge on channel walls. This allows us to exclude the first-order effects of steric and Coulomb-based exclusion. We show that the studied two-dimensional Å-scale capillaries can distinguish between same-charge ions with similar hydrated diameters. The selectivity is attributed to different positions occupied by ions within the layered structure of nanoconfined water, which depend on the ion-core size and differ for anions and cations. The revealed mechanism points at possibilities of ion separation beyond the simple steric sieving.**

**Introduction**

Ions in aqueous solutions acquire hydration shells due to the electric field of a charged ion, which forces dipolar molecules of water to rearrange around and screen the field. Accordingly, two types of diameters (Fig. 1) are attributable to ions in water: i) ionic diameter $D_I$ given by the ion's core and ii) hydrated diameter $D_H$ that includes polarized water molecules around ions[1,2]. To notably alter ion permeation, a confinement comparable with $D_H$ is essential. Otherwise, ions can move through capillaries like in bulk solutions, without experiencing discernable barriers. For common salts, $D_H$ is generally smaller than 10 Å[1,3]. Biological protein channels (e.g., potassium, calcium, sodium, or chloride channels) have angstrom-scale sizes[4-6] and can transport ions across cell membranes with selectivity factors up to ~$10^2$–$10^4$. Such a high selectivity is possible due to a combination of several mechanisms including atomic-scale confinement, Coulomb repulsion, and subtle interactions between ions and channel walls. The well-known example is K$^+$ channel[7] that allow nearly perfect permeation of potassium ions whereas transport of Na$^+$ is practically forbidden, despite the latter ions having similar $D_H$ and even smaller $D_I$. This selectivity is attributed to the seamless ("snugly") fitting of dehydrated K$^+$ ions between functional groups covering the protein channel, which reduces the energy barrier associated with hydration-dehydration processes required for the ion's transport through[5,8-10].

Trying to mimic the nature and achieve high selectivities, artificial channels based on carbon nanotubes, graphene, MoS$_2$, graphene oxide, zeolites, covalent and metal-organic frameworks, etc. have been intensively studied over the past two decades[11-16]. Many factors controlling ion transport under strong confinement were also uncovered, including the roles played by hydration shells' size and polarization[13,17-19]. Certain nanoporous materials were found to exhibit high selectivity with respect to one or another ion[10,20,21], but generally it has proven difficult to achieve and control the selectivity, especially for the case of same-charge ions with similar $D_H$, which would have to rely on mechanisms beyond the simple steric or Coulomb exclusion. More recently, van der Waals assembly of two-dimensional (2D) crystals has allowed controllable fabrication of angstrom-scale channels (Å-channels) and studies of ion transport through them[22]. The channels exhibit pronounced steric ion selectivity such that cations' permeation was found to decrease by more than an order of magnitude[23,24] if $D_H$ exceeded the channel height $h$ just by < 50%. The reduction of the ion mobility is due to steric hindrance such that ions larger than $h$ cannot enter inside the channels without partial shedding and rearranging of their hydration shells. However, it remains unclear what other factors govern the ion transport if $D_H$ is not the key factor. How do ions behave under the confinement and how are water molecules rearranged around ions inside the channels? In this report, we show that Å-channels can provide selectivity factors larger than 10 for same-charge ions with similar $D_H$ and, in some cases, can even distinguish between ions with similar $D_H$ and $D_I$ We focus on monovalent ions with hydrated diameters of about 7 Å, which are all close in size to the precisely controlled $h ≈ 6.8$ Å of our Å-channels. This allows us to exclude the first-order effects of steric and Coulomb exclusion and investigate factors beyond them.

The second-order effects are still surprisingly strong and, although subtle in nature, provide important insights for future design of ion selective channels.

**Results and discussion**

The studied channels were fabricated following the recipe described elsewhere[25]. Briefly, bilayer graphene was patterned into narrow strips and placed between two relatively thick (≲ 200 nm) crystals of graphite, hexagonal boron nitride (hBN) and molybdenum sulfide (MoS$_2$). The crystals served as top and bottom walls of the resulting 2D channels while the graphene strips (bilayer thickness is ~6.8 Å) defined the channel height, $h$ (Fig. 1a right panel and Supplementary Figs. 1 & 2). The tri-crystal assembly was kept together by van der Waals forces[25]. Bilayer spacers were chosen because $h \approx 6.8$ Å provides a close match to $D_H$ of typical inorganic ions, like those in seawater. This $h$ allows rather strong 2D confinement but, at the same time, does not forbid transport of hydrated ions[23], as is the case of Å-channels with thinner, monolayer spacers[24]. Each channel was ~120 nm wide ($w$) and several microns long. To increase experimental accuracy, we usually used a few hundred ($N$) channels acting in parallel[23]. The multichannel devices were assembled on top of a silicon nitride membrane with a prefabricated rectangular hole, which separated two containers filled with salt solutions and provided the liquids access to the opposite sides of Å-channels (Fig. 1a). As the channel walls were basal planes of mechanically exfoliated 2D crystals, their surfaces held little electric charge and exhibited weak interaction with both ions and water, as reported previously[23]. To measure ionic conductivity through Å-channels, we used both Ag/AgCl and reference electrodes and recorded current-voltage ($I$-$V$) characteristics using voltages up to typically ± 100 mV (Fig. 1; see Supplementary section 2). The devices were wet using isopropanol-water solutions, then the aqueous salt solutions were filled carefully by slow rinsing steps. This served to remove small air bubbles that otherwise could block channel entrances. Prior to detailed studies, we routinely cycled the applied voltage $V$ for several hours, which allowed the removal of hydrocarbon contamination that could often block some Å-channels. As a reference, we used devices fabricated in the same manner as Å-channels but without placing any spacers. Those exhibited a leakage conductance of only ~10-20 pS. For ion conductance measurements, we used potassium salts with different anions F$^-$, Cl$^-$, I$^-$ and ClO$_4^-$ and chloride salts with cations Cs$^+$, Na$^+$, and Li$^+$ (see Fig. 1a left panel). This selection provided anions and cations with close $D_H$ ranging from 6.0 to 7.6 Å, and closely matching $h \approx$ 6.8 Å (ions larger or smaller than the $h$ by within ~12%) but rather a different $D_I$ (varying by a factor of > 2, from 1.9 to 4.5 Å), as listed in Supplementary Table 1. Typical $I$-$V$ curves for the ionic solutions are shown in Fig. 1b for the concentration $C \approx 0.1$ M in both containers. The curves were linear allowing us to define the ionic conductance $G = I/V$. Repeating the measurements for devices of different lengths $L$ from 1 μm to 8 μm, we found that the channel resistance increased linearly with $L$ (Fig. 1b, inset). This shows that capillary-entry effects were small so that we could evaluate conductivities of the confined salt solutions using $\sigma_A = G \times L/(Nwh)$. Our Å-channel devices with walls made from graphite, hBN and MoS$_2$ exhibited the same $\sigma_A$ within experimental scatter (Fig. 2a and Supplementary Fig. 3). A clear trend KF > KI > CsCl ≈ KCl > NaCl > LiCl > KClO$_4$ is seen for $\sigma_A$ in Figs. 1b & 2a. This behavior differs qualitatively from that of the same bulk solutions[26] that all show similar conductivities (Fig. 2a).

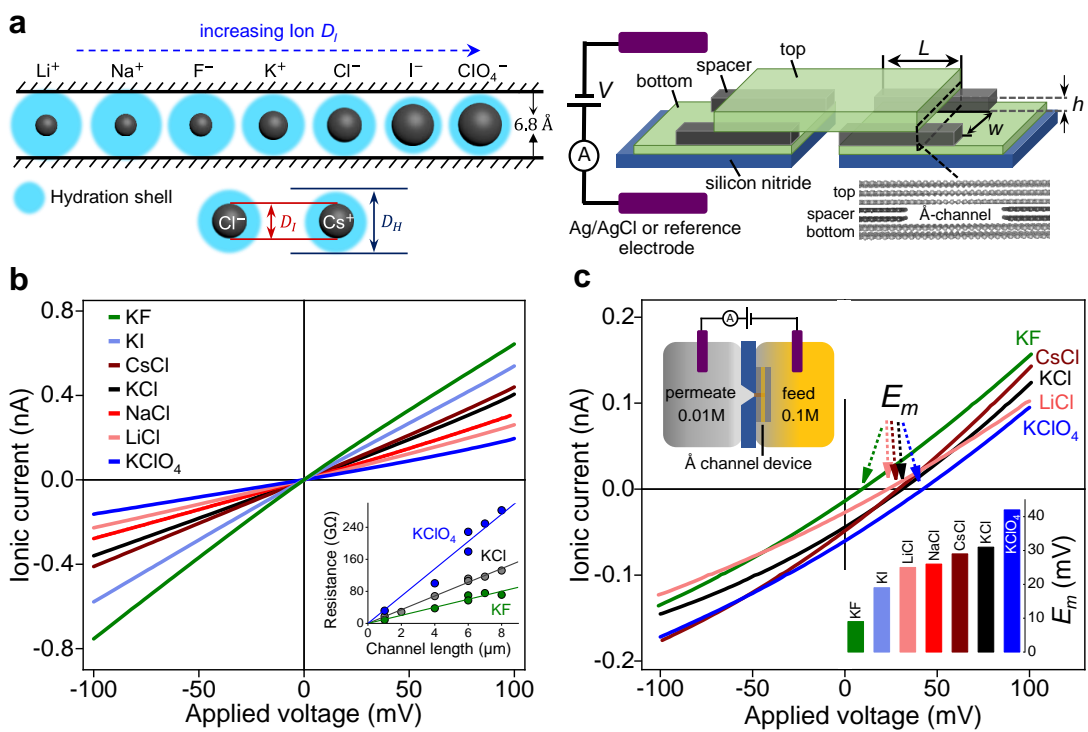

**Fig. 1| Conductivity of salt solutions under Å-scale confinement.** (**a**) Left side panel: cartoons of the studied ions which have similar hydrated diameters ($D_H$) but different ionic diameters ($D_I$), with their size shown relative to the 6.8 Å channel height. The $D_I$ is increasing from left to right. Right side panel: Schematics of Å-channels device and measurement setup. *L*, *w*, and *h* denote the length, width and height of the channel. Inset, a cross-sectional view of an Å-channel, which is made from three layers, top, spacer and bottom as indicated. (**b**) *I-V* characteristics for various salts (color-coded) measured for a hBN channels device with $w \approx$ 120 nm, $h \approx$ 6.8 Å, $N \approx$ 400, and $L \approx$ 6 μm; $C$ = 0.1 M. Inset, resistance per channel as a function of *L* for 6 devices using KClO$_4$, KCl and KF solutions. (**c**) Color-coded *I-V* curves using 0.01 M and 0.1 M salt solutions in the opposite containers, as shown schematically in the top inset. Bottom inset, $E_m$ measured for the studied salts.

To gain further information about ion transport under Å-scale confinement, we performed drift-diffusion experiments[23,27,28] (Fig. 1c, top inset) in which the ionic current is driven by both *V* and concentration ratio ($\Delta C = \frac{C_{high}}{C_{low}}$). Examples of such measurements are shown in Fig. 1c for $\Delta C$ = 10. The zero-current potential $E_m$ represents the voltage required to balance ionic currents arising due to different mobilities $\mu$ of cations and anions ($\mu^+$ and $\mu^-$, respectively) (Fig. 1c bottom inset). For example, large positive $E_m$ as in the case of KCl and KClO$_4$ in Å-channels, implies that the mobility of $K^+$ is considerably higher than that of $Cl^-$ and $ClO_4^-$, in contrast to the case of bulk[26] KCl and $KClO_4$ where the ions have same $\mu$. The measured $E_m$ values can readily be translated into the mobility ratio $\mu^+/\mu^-$ using the Henderson equation[23,29]

$$\frac{\mu^+}{\mu^-} = \frac{\ln(\Delta C) + eE_m/k_B T}{\ln(\Delta C) - eE_m/k_B T} \quad \text{Eq. (1)}$$

which is simplified here for our case of monovalent ions ($k_B$ is the Boltzmann constant, $e$ is the electron charge and $T$ = 295 ± 2 K is the temperature in our experiments). The found mobility ratios are shown in Fig. 2b. As a control, we measured drift-diffusion curves for micrometer-sized holes in a silicon nitride membrane using both Ag/AgCl and reference electrodes (Supplementary Fig. 4 and Supplementary Table 2) and found $\mu^+/\mu^-$ close to the values known for bulk solutions (Fig. 2b)[26]. Taking into account that conductivity is given by $\sigma = eN_A C (\mu^+ + \mu^-)$ where $N_A$ is the Avogadro number, the measurements of both $\sigma_A$ and $E_m$ allow us to find $\mu^+$ and $\mu^-$ separately. We have considered $\sigma_A$ of 0.1 M salt solutions to calculate the individual ion mobility, assuming that the ion concentration inside the channel and the bulk are similar (Supplementary section 4).

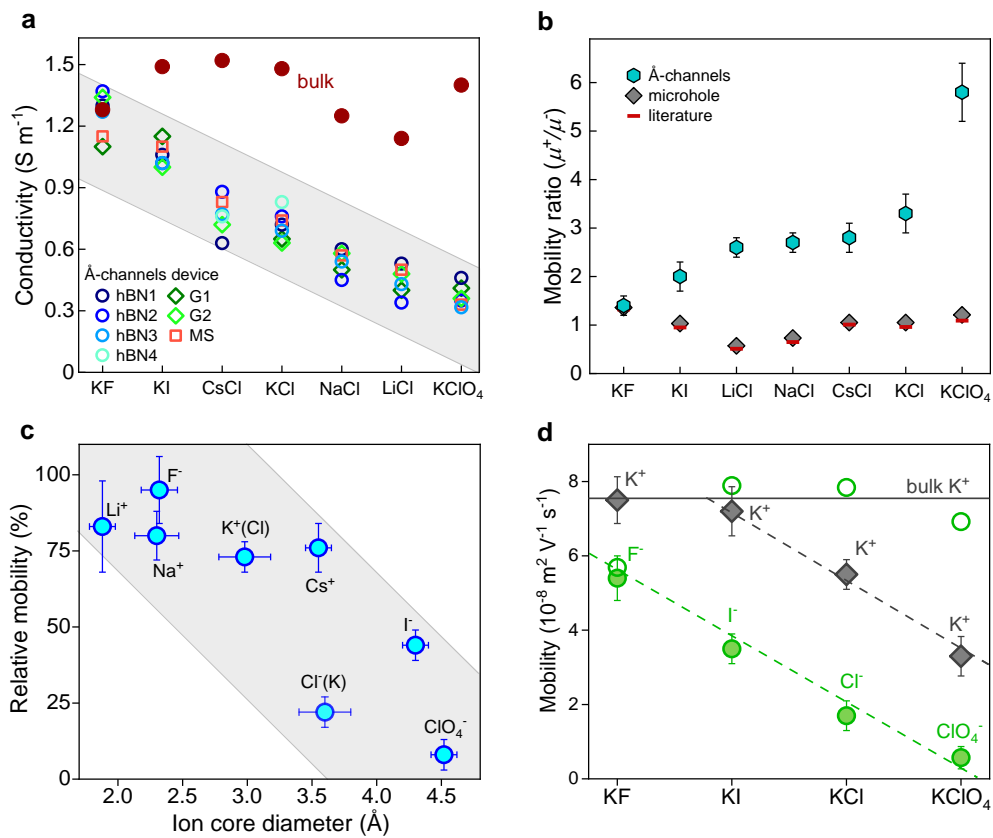

**Fig. 2| Effect of ions' core size on their mobility under Å-scale confinement.** (**a**) Conductivity ($\sigma_A$) for the studied salt solutions in Å-channels with hBN (circles), graphite (diamonds) and $MoS_2$ (square) walls (G1, G2 refer to graphite devices, and MS to a $MoS_2$ device). The corresponding bulk conductivities are shown by filled circles. (**b**) The measured mobility ratio

($\mu^+/\mu^-$) of different salts using Å-channels (hexagons) and microholes (diamonds). Horizontal lines represent literature values for the bulk solutions. (**c**) Relative mobilities of Å-confined ions (with respect to the bulk values) as a function of ionic diameter ($D_I$). K$^+$ and Cl$^-$ mobilities are indicated here for KCl solution (**d**) Mobilities of various anions (filled circles) with the same counterion K$^+$ (filled diamonds) measured for different salt solutions under the Å-scale confinement. The bulk mobilities for the corresponding anions are shown by open circles, whereas the bulk K$^+$ mobility is indicated by the solid line. Dashed lines in (d), guide to the eye. The horizontal error bars in (c) indicate the spread in the $D_I$ values from the literature[1,2]. Vertical error bars in b, c, and d, the average ± standard deviation (SD) in mobilities obtained from three devices. Shaded areas in a and c are guide to the eye.

From our measurements, the extracted individual ion mobilities in the Å-scale confinement are notably suppressed for some of the ions (Fig. 2c) . If a hydration diameter notably exceeds $h$, diffusion of confined ions becomes strongly suppressed and they can even be excluded from the smallest Å-channels[23,24]. However, for our case of $h \approx D_H$, such steric effects were found to be small[23]. For example, Li$^+$, Na$^+$, and F$^-$ have largest $D_H$ among the studied ions (up to ~10% larger than $h$) but their mobilities are suppressed by no more than 25% with respect to the bulk, in contrast to ions with smaller $D_H$ (I$^-$ and ClO$_4^-$) which exhibit much stronger suppression (~12 times for ClO$_4^-$). Similarly, there is no preferential selectivity between anions and cations (cf. mobilities of confined F$^-$ and Na$^+$), which rules out a possible influence of residual charges on Å-channels' walls[23]. On the other hand, if we plot the found mobility suppression as a function of $D_I$, a clear tendency emerges: the larger the ion core the stronger the confinement-induced suppression of $\mu$ tends to be (Fig. 2c). Using potassium bearing salts with different anions, we have measured mobility of K$^+$ in the presence of different anions (F$^-$, I$^-$, Cl$^-$ and ClO$_4^-$) that in turn exhibit rather different mobilities within our Å-channels. Fig. 2d and Supplementary Fig. 10 show a strong effect of anions on K$^+$ transport. If anion's mobility is suppressed strongly by the confinement, the cation also shows a notable suppression. For instance, in a KF solution, F$^-$ exhibits little change in $\mu$ under the Å-scale confinement. Accordingly, mobility of K$^+$ is also found little affected by the confinement. In contrast, for the case of KClO$_4$ solutions, in which the ClO$_4^-$ mobility was suppressed by ~10 times with respect to the bulk, the K$^+$ mobility is also notably suppressed (by a factor of ~2.2). Qualitatively, this can be understood as the hindrance to K$^+$ diffusion by slowly-moving counterions that occupy a considerable amount of space within the 2D channels (Supplementary Fig. 10).

To understand the mechanism behind the unexpected beyond steric selectivity, we have carried out molecular dynamics (MD) simulations (Supplementary Figs. 5 to 9). Our 6.8 Å channels can accommodate only two monolayers of water[30] and, therefore, ions inside acquire rather distorted and partially dehydrated shells (Fig. 3a). The probabilities of finding the studied ions at different positions $z$ across the Å-channel (in the vertical direction) are shown in Fig. 3b and Supplementary Fig. 5. Some ions (F$^-$, K$^+$ and Na$^+$) tend to reside close to the channel's center (within $z = \pm 1$, with $z = 0$ at the centre), whereas others (Cl$^-$, ClO$_4^-$, Cs$^+$, Li$^+$ and I$^-$) stay adjacent to the walls. These MD observations are supported by our free energy calculations (Supplementary Fig. 6), which show that, along the vertical direction in the channel, the ions' energy minima appear close to their MD probability maxima. To quantify the positional differences, Fig. 3c plots probabilities of finding the ions near the Å-channel walls as a function of $D_I$. The figure reveals that, as the ion core size increases, ions gradually shift their preferable location from the channel center towards the walls. For example, small ions such as $K^+$ and $F^-$ have core diameters notably smaller than the ~4 Å thickness of bilayer water (Supplementary Fig. 8), which allows the ions to fit in without a major disruption of the first hydration shells (Fig. 3a). On the other hand, ions with large cores cannot fit between the water layers and are pushed towards the walls so that their first hydration shells are partially shaved off ($Cl^-$ and $Cs^+$ in Fig. 3a). An exception is $Li^+$ which can be accommodated within a monolayer of water because of its particularly small $D_I$ [anomalous positioning of $Li^+$ ion near-surface water was also noted previously[31]]. Based on the above MD analysis, it is instructive to replot the found suppression of $\mu$ as a function of ions' probability to reside near the channel walls (Supplementary Fig. 10). It is clear that mobilities of ions staying close to the walls are notably suppressed as compared to ions in the channel center.

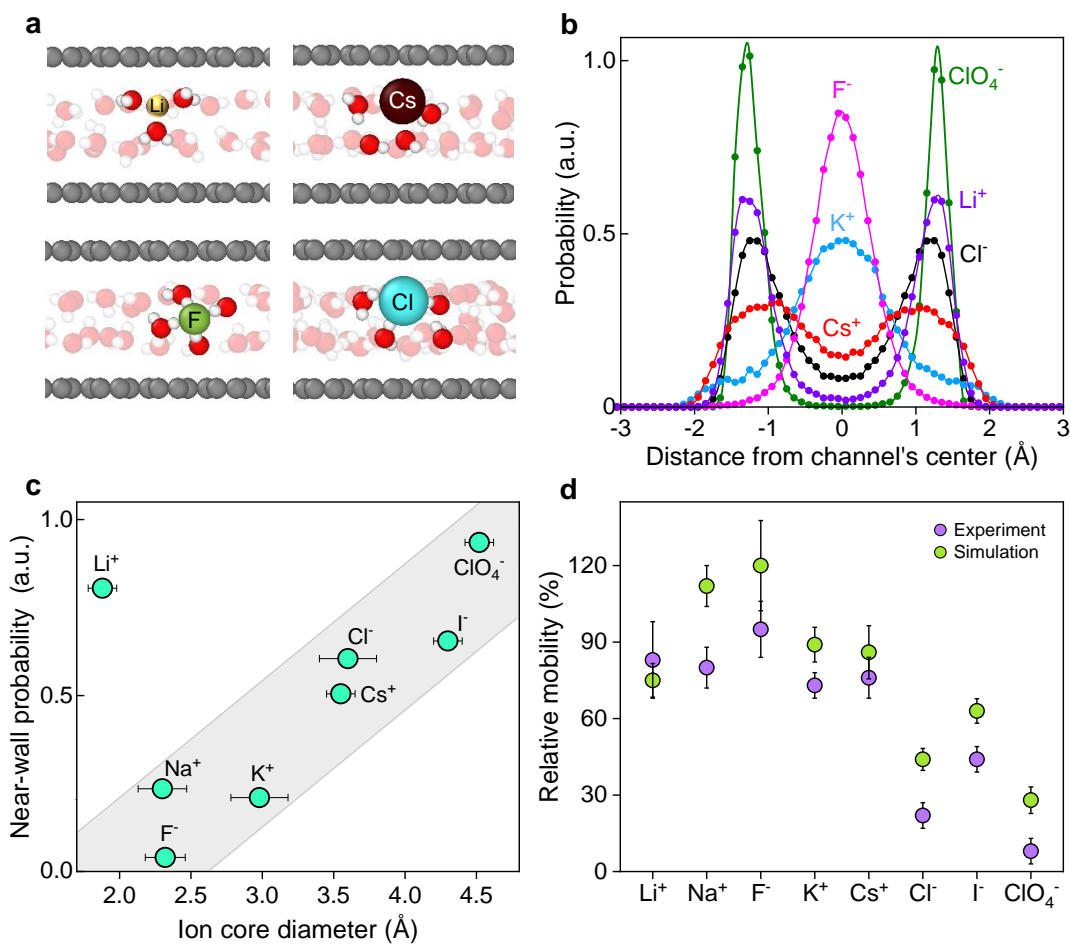

**Fig. 3| Ion positioning in 2D water and mobility reduction.** (**a**) Snapshots from MD simulations showing local arrangements of water molecules around ions in Å-channels. (**b**)

Probability of finding various ions at different distances $z$ from the channel center. (**c**) Probability of finding ions close to the channel walls at $|z|>1$ Å. The shaded area is a guide to the eye. Horizontal error bars indicate the spread in the $D_l \pm$ standard deviation (SD) values from the literature[1,2] (**d**) Mobility of ions confined in Å-channels relatively to the bulk mobility, as measured experimentally and simulated. In (**d**) K$^+$ and Cl$^-$ mobilities are from KCl salt. In **d**, the vertical error bars for experimental data are from ±SD of the ion mobility from three devices, and for simulations data, the error was obtained from ±SD of over ten repeat simulations using different seed points for each simulation.

To relate the found differences in ions' location within 2D water to changes in their $\mu$, we have calculated their mean square displacement as a function of time and found the diffusion coefficients $D$ (see the Supplementary Section 6). The Einstein relation was then used to evaluate $\mu = \frac{e}{k_\mathrm{B} T} D$. The MD results are plotted in Fig. 3d and show good agreement with the experiment. Furthermore, our simulations using different wall materials found no notable differences in either ion positions or mobilities (Supplementary Fig. 7), again in agreement with the experiment. The reason why ions residing near the walls exhibit suppressed mobilities is the strongly distorted hydration shells, which destroy translational symmetry within 2D water. We studied the orientations of water molecules around an ion residing in the centre (K$^+$) and another near the wall (Cl$^-$), which showed large difference in the angular orientations when confined in our Å-channels, indicating a distortion of the hydration shells (Supplementary Figs. 11 and 12). As the near-wall ions diffuse along Å-channels, they constantly have to change the arrangements of water molecules around them, which can also be viewed as enlarged "fur coats" the ions carry around[30]. The case of Li$^+$ illustrates the importance of such shell distortions. The small ion sits near the walls but does not strongly distort the water structure nor exhibits a large reduction in $\mu$. Note that its mobility under the Å-scale confinement falls into the general sequence presented in Fig. 3d, despite Li$^+$ being an outlier in Fig. 3c because of its near-wall location.

Going further than the limitations of classical MD simulations, we investigate ion-water interactions using ab-initio simulations (Supplementary Figs. 13 and 14), which reveal the distortion of ions' hydration shells and a reduction in the coordination number of water molecules around an ion[32] (Supplementary Figs. 13 and 14). The hydration energy also decreased when the ions enter the channel (Supplementary Table 5). Raman or vibrational spectroscopy and transmission electron microscopy measurements of ions and the surrounding water in confined channels could be interesting, however require much further dedicated efforts in improving the signal from two layers of water.

In addition to the importance of the ion core size, analysis of the experimental data in Fig. 2c reveals two other factors playing roles in the suppression of ion mobilities under strong confinement. One is the ion charge; its influence is clearly seen in the case of Cl$^-$ and Cs$^+$ which have practically the same $D_H$ and $D_l$ but exhibit 3 times different $\mu$. With little surface charge on the channel walls, the effect can be attributed to different polarization of ion hydration shells. This leads not only to slightly different probabilities for the two ions to be near the channel walls (Figs. 3b & 3c) but, more importantly, to different interaction of ions with channel walls. Indeed, 2D materials like graphene, hBN and $MoS_2$ have a stronger affinity towards $OH^-$ groups[33,34] than $H^+$ and, because anion's hydration shells have $OH^-$ groups directed towards the confining surfaces, $\mu^-$ is expected to be suppressed stronger than $\mu^+$. The difference between mobilities of $Cs^+$ and $Cl^-$ is also captured by our MD calculations (Fig. 3d). The second factor is a trend towards lower $\mu$ for ions if their counterions are poorly mobile under confinement (Fig. 2d). Indeed, $K^+$ exhibits twice different mobilities if $F^-$ or $ClO_4^-$ is the counterion (Fig. 2d). Such influence does not happen in the bulk, and we attribute the counterion effect to the reduced space available for ion diffusion in 2D if counterions become immobile. For example, $ClO_4^-$ is almost static with respect to rapidly diffusing $K^+$ and the space $ClO_4^-$ occupies limits possible trajectories for the former ion. The spatial restrictions are known to become increasingly important with decreasing dimensionality. Alternatively, Coulomb attraction between cations and anions that is predicted to be enhanced in 2D water can also play a role[35]. Further experimental and theoretical studies are required to clarify the origins of the unexpected influence of counterions' mobility.

To conclude, our work shows that strong geometrical confinement can lead to notable selectivity between similar-size anions. Opposite charge ions with practically the same hydrated and ion-core diameters (e.g., $Cs^+$ and $Cl^-$) can also be distinguished. The demonstrated selectivity relies on difference in ions' positions inside Å-scale channels, which depend on a subtle combination of ion's parameters and affect their mobilities. The revealed selectivity mechanism resembles the way governing ion selectivity in biological channels in which snugly fitting of ions between the walls plays a crucial role. Spectroscopic and electron microscopy techniques can be applied in future to investigate the inferred structure of confined water and hydration shells. 2D channels with functionalized walls should allow higher selectivities and may offer a venue towards development of designer sieves to filter out chosen ions.

**Acknowledgements**

B.R. acknowledges the funding from Royal Society university research fellowship URF\R1\180127 and Philip Leverhulme Prize PLP-2021-262. B.R., S.G., A.I., Y.Y., acknowledge funding from the European Union's H2020 Framework Programme/ERC Starting Grant 852674 – AngstroCAP, RS enhancement award RF\ERE\210016, EPSRC New horizons grant EP/X019225/1. A.K. acknowledges Ramsay Memorial Fellowship, and Royal Society International Exchanges grant IES\R1\201028, and EPSRC new horizons grant EP/V048112/1. F.C.W. acknowledges the Youth Innovation Promotion Association CAS (2020449) and Hefei Advanced Computing Center. H.J and M. N-A would like to acknowledge high-performance computing support from Shahid Rajaee TT-University.

**Author contributions**

B.R., designed and directed the project. A.K., Y.Y., A.B., carried out sample fabrication of several Å-channel devices. S.G., A.I. performed ion conductance measurements and their analysis. S.G. conducted diffusion measurements and their analysis. A.K., M.V.S.M carried out sample characterization. H.J., N.H., M.N.A., Y.L., H.P., F.W provided theoretical simulations. B.R., S.G., A.I., A.K., wrote the manuscript with inputs from M.N.A., F.W. All authors contributed to discussions.

**Competing interests**

The authors declare no competing interests.

# Supplementary Information for

# Beyond steric selectivity of ions using angstrom-scale capillaries

Solleti Goutham[1,2], Ashok Keerthi [2,3], Abdulghani Ismail[1,2], Ankit Bhardwaj[1,2], Hossein Jalali,[4] Yi You[1,2], Yiheng Li,[5] Nasim Hassani,[4] Haoke Peng,[5] Marcos Vinicius Surmani Martins[1,2], Fengchao Wang,[5] Mehdi Neek-Amal,[4,6] Boya Radha[1,2]*

[1]Department of Physics and Astronomy, School of Natural Sciences, The University of Manchester, Manchester M13 9PL, United Kingdom
[2]National Graphene Institute, The University of Manchester, Manchester M13 9PL, United Kingdom
[3]Department of Chemistry, School of Natural Sciences, The University of Manchester, Manchester M13 9PL, United Kingdom
[4]Department of Physics, Shahid Rajaee Teacher Training University, 16875-163 Lavizan, Tehran, Iran
[5]Chinese Academy of Sciences Key Laboratory of Mechanical Behavior and Design of Materials, Department of Modern Mechanics, University of Science and Technology of China, Hefei, 230027, China.
[6]Department of Physics, University of Antwerp, Groenenborgerlaan 171, Antwerp B-2020, Belgium

*Correspondence to: radha.boya@manchester.ac.uk

**Contents**

1. Device fabrication
2. Electrical measurements
3. Control drift-diffusion experiments using microholes
4. Individual ion mobility calculation in experiments
5. Molecular dynamics simulations
6. Mobility and diffusion coefficient from simulations
7. Effects of walls and counterions
8. Angular orientations of water molecules around ions
9. *Ab initio* simulations for determining hydration shell and energy
10. Supplementary references

**Supplementary Section 1. Device fabrication**

The angstrom (Å)-channels were fabricated following the recipe reported previously[1,2]. The channels could be viewed as van der Waals heterostructures made of three atomically flat crystals. Relatively thick hBN, graphite or $MoS_2$ crystals served as top and bottom walls of the Å-channels whereas strips of bilayer graphene were placed in between and served as spacers. All the 2D crystals were prepared by mechanical exfoliation of bulk graphite, hBN or $MoS_2$ on an oxidized silicon wafer (~ 290 nm $SiO_2$ on Si).

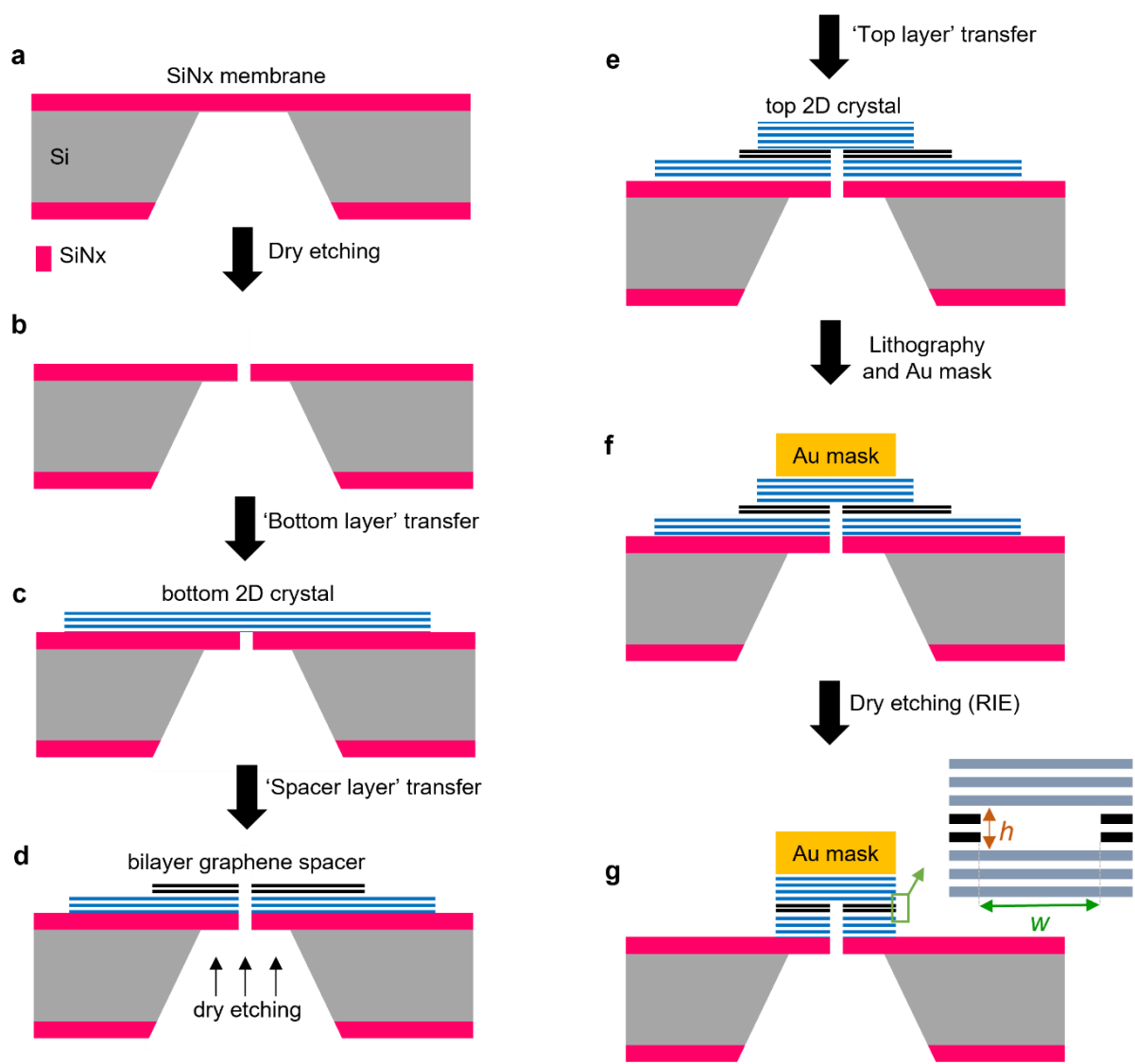

**Supplementary Fig. 1| Fabrication process flow.** The sequence of fabrication steps is indicated by the arrows. (**a**) Freestanding silicon nitride ($SiN_x$) membrane (100 × 100 μm²) was prepared by wet etching. (**b**) Rectangular aperture (typically, 3 × 26 μm²) was made in the membrane using photolithography and dry etching. (**c**) Bottom 2D crystal was placed on top of the aperture. (**d**) Strips of bilayer graphene were transferred onto the bottom crystal. Then the two-layer assembly was dry etched from the back with the $SiN_x$ aperture serving as a protective mask. (**e**) Another 2D crystal (top layer) was transferred to cover the entire etched micro-hole. (**f**) A gold stripe was deposited on top of the trilayer assembly. (**g**) This stripe was used as a mask for subsequent dry etching. The inset shows schematically a cross-section of the resulting Å-channels.

First, we used a commercial Si wafer covered with a ~500 nm thick layer of silicon nitride ($SiN_x$) to prepare a freestanding $SiN_x$ membrane (Supplementary Fig. 1a). A rectangular aperture (microhole) was dry etched through the freestanding membrane (Supplementary Fig. 1b). The resulting $SiN_x$ membrane was thoroughly cleaned using acetone and isopropanol (IPA) followed by a 10-minute exposure to an oxygen plasma. Then a thin (10 - 20 nm) crystal of graphite, hBN or $MoS_2$ was transferred onto the $SiN_x$ membrane to cover the aperture (Supplementary Fig. 1c). It was important to have the $SiN_x$ surface clean and the bottom crystal thin, which ensured their strong adhesion. A bilayer graphene crystal (height $h \approx 6.8$ Å) was prepared on an oxidized Si wafer and patterned into many parallel strips (width $w$ of $120 \pm 10$ nm) which were separated by the same distance. The patterning was done by e-beam lithography, and poly(methyl methacrylate) (PMMA) served as a mask for etching the bilayer graphene in an oxygen plasma. The strips were then cleaned using mild sonication in acetone to remove the remaining PMMA. They were transferred onto the bottom crystal (graphite, hBN or $MoS_2$) and aligned perpendicular to the rectangular microhole (Supplementary Fig. 1c). After that, we used a reactive ion etch (RIE) plasma (oxygen for graphite, and a mixture of $CHF_3$ and oxygen for hBN and $MoS_2$) to etch from the backside of the membrane, which extended the rectangular microhole into the two-layer assembly (Supplementary Fig. 1d). The whole stack was annealed at 400 °C in a hydrogen-argon atmosphere (10% $H_2$ in Ar) for 5 hours. A relatively thick (~ 150 - 200 nm) top crystal (hBN, graphite or $MoS_2$) was then transferred onto the assembly to cover the entire aperture (Supplementary Figs. 1e & 2a). The thickness of the top crystal was important to avoid its sagging into Å-channels[1]. To accurately define the channel length $L$, we deposited a wide gold stripe (5 nm Cr/ 50 nm Au) onto the top crystals by using photolithography (Supplementary Figs. 1f & 2b). The stripe served as a protective mask for dry etching so that the uncovered parts of the trilayer assembly were removed (Supplementary Figs. 1g & 2c). This step not only defined the channel length but also opened Å-channels' entries that could otherwise be blocked by sagged thin near-edge parts of the top crystal and/or by hydrocarbon contamination. The resulting devices were again annealed in $H_2$-Ar at 400 °C for 5 hours. To avoid contamination from air[3], the devices were stored in deionized (DI) water or 100% humid environment, before and between the electrical measurements. Optical images of our pre-final and final devices are shown in Supplementary Fig. 2.

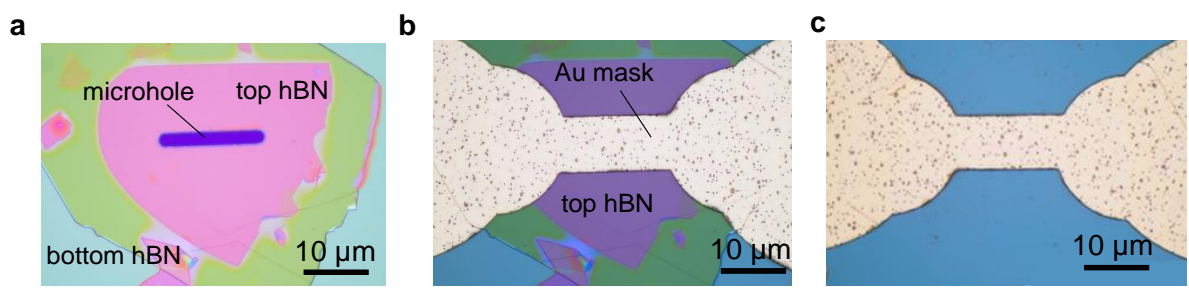

**Supplementary Fig. 2| Optical micrographs of Å-channel devices.** (**a**) Top-view of the trilayer assembly (hBN-graphene spacers-hBN) before depositing a gold mask. The graphene spacer is not clearly visible as it is very thin (height, ~0.7 nm). The rectangular microhole in the $SiN_x$ membrane is seen in dark blue. Top view of the same device (**b**) after the gold mask was deposited and (**c**) after etching away the exposed areas of the 2D crystals.

**Supplementary Table 1|** Ion core and hydrated diameters for the studied ions as reported in the literature.

| Ion | Ionic diameter (Å) | Hydrated diameter (Å) |
|---|---|---|
| $F^-$ | 2.32 ± 0.14 (ref 4-6) | 7.04 (ref 7) |
| $Cl^-$ | 3.6 ± 0.2 (ref 5-7) | 6.64 (ref 7) |
| $I^-$ | 4.3 ± 0.1 (ref 7,8) | 6.00 (ref 4) |
| $ClO_4^-$ | 4.52 ± 0.1 (ref 5,9) | 6.76 (ref 7) |
| $K^+$ | 2.98 ± 0.2 (ref 6,10) | 6.62 (ref 7) |
| $Cs^+$ | 3.55 ± 0.1 (ref 6,7,11) | 6.58 (ref 7) |
| $Na^+$ | 2.30 ± 0.17 (ref 6,8) | 7.16 (ref 7) |
| $Li^+$ | 1.88 ± 0.1 (ref 6,11) | 7.64 (ref 7) |

**Supplementary Section 2. Electrical measurements**

We used a custom-made electrochemical cell machined from polyether ether ketone (PEEK). The cell consisted of two small reservoirs and allowed our $SiN_x$ wafers with Å-channels to be clamped in between using two O-rings (Figs. 1a & 1c of the main text). Each reservoir could hold ~2.5 mL of an electrolyte solution and had slots for inserting electrodes into it. Before mounting our devices, the cell and O-rings were thoroughly washed in DI water and dried under a nitrogen gas flow. Once a device was in place, we filled the reservoirs with liquid solutions that were prepared using *Milli-Q* DI water (resistivity of ~18 MOhm cm). Liquids were poured inside the reservoirs slowly using a 3 mL pipette. To ensure that no air bubbles were blocking entries/exits of Å-channels, we first wet the devices using isopropanol (IPA)-water mixtures, gradually increasing the IPA concentration from 10% to 50%, and then rinsing the cell with pure DI water. After this, the cell could be filled with salt solutions that were freshly prepared before each set of measurements.

The *I-V* characteristics were measured using the standard Ag/AgCl or reference (*CH Instruments, USA*) electrodes and *Keithley* 2636B Source Meter. Data were acquired using *LabVIEW*. Typically, the applied voltage *V* was swept between ± 100 mV with a rate of 2 mV per sec. All measurements were done at room temperature (295 ± 2 K). In control experiments, we prepared 'dummy' devices following the same trilayer assembly procedures but without etching bilayer graphene into strips. The conductance of such devices was typically ~10-20 pS in 100 mM KCl solutions, two-three orders of magnitude smaller than for our devices with Å-channels.

From the linear slopes of the measured *I-V* characteristics, we could extract the conductance *G* of confined solutions. First, we used 0.1 M KCl solutions to compare the extracted *G* with the values expected for the given channel dimensions (*N*, *L* and *w*) and the known bulk conductivity[2,12]. KCl solutions were used because previously it was shown that their conductivity within 6.8 Å channels was only slightly (by a factor of ~1.5) suppressed with respect to the bulk[2]. If much-higher-than-expected *G* was observed, this indicated delamination of the van der Waals assembly, and such devices were discarded. Alternatively, if the measured *G* was considerably (upto two orders of magnitude) lower than the expected value for a 0.1 M KCl solution, this meant that channels were blocked. Typically, only half of our devices could show proper conductance. The other half had to be discarded because of delamination or channels' blockage. Furthermore, only a few of our devices sustained the entire set of planned measurements using several salts. Other operational devices could suddenly exhibit a large increase in *G* during measurements, which again indicated their delamination. To ensure that no delamination occurred during the measurements of different salt solutions, each device was tested at the end with 100 mM KCl. Only those devices that exhibited no change in *G* with respect to their original conductance have been presented in this report. There were seven such devices. In addition to Fig. 1b of the main text (for Å-channels with hBN walls), further examples of *I-V* characteristics with graphite and MoS$_2$ channel devices are shown in Supplementary Fig. 3.

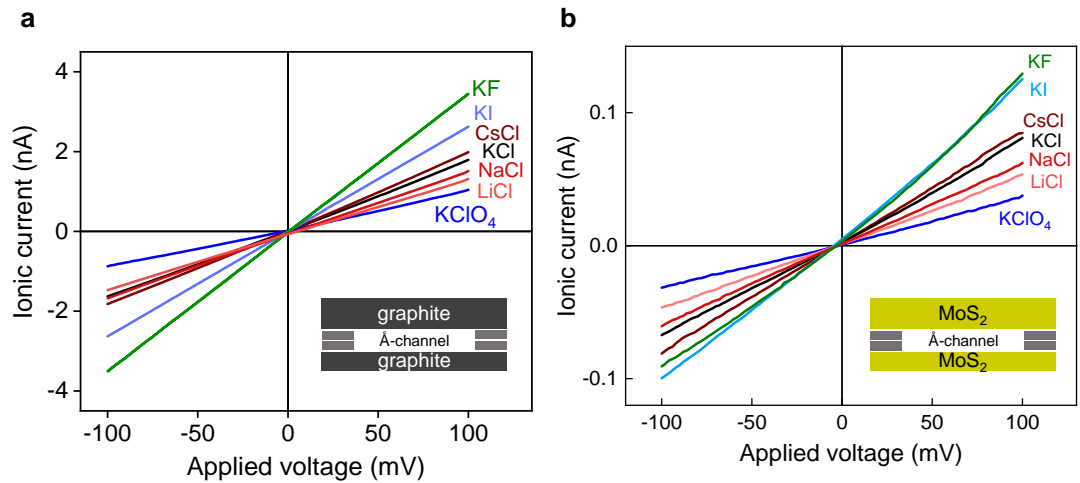

**Supplementary Fig. 3| Ion transport measurements with graphite and MoS$_2$ walls capillaries.** *I-V* characteristics measured for a device (*h* = 6.8 Å) with bottom and top (**a**) graphite walls; $w \approx 120$ nm, $L \approx 1$ μm and $N \approx 300$; (**b**) MoS$_2$ walls; $w \approx 120$ nm, $L \approx 7$ μm and $N \approx 88$. Various salt solutions (color-coded) are measured for concentrations of 0.1 M in both reservoirs. The insets in both (a) and (b) show the cross-sectional schematic of graphite channel and MoS$_2$ channel with bilayer graphene spacers, respectively.

**Supplementary Section 3. Control drift-diffusion experiments using microhole**

A rectangular microhole (3 × 30 μm$^2$) made in the SiN$_x$ membrane was used as a reference device to compare ion mobilities in the bulk with those under Å-scale confinement. The microhole device was mounted in the same PEEK cell, and the standard Ag/AgCl or reference electrodes were used for drift-diffusion measurements. The two reservoirs were filled with various salt solutions having a Δ$C$ = 10 (10 mM and 100 mM in the opposite containers; top inset of Supplementary Fig. 4). The zero-current potential was defined from the intersection of *I-V* characteristics with the x-axis (Supplementary Fig. 4). Its non-zero value reflected a difference between cation and anion mobilities as explained in the main text[13,14]. The higher the difference in mobilities the further the zero-current potential was from the origin.

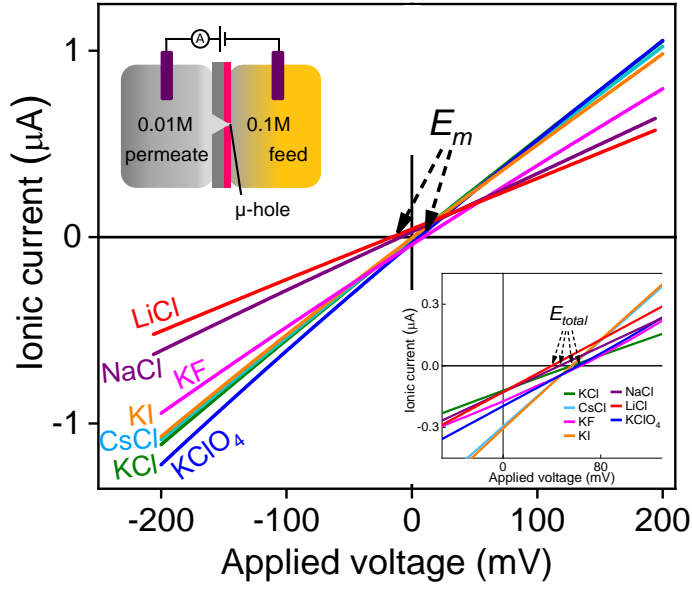

**Supplementary Fig. 4| Drift diffusion measurements using a microhole device**. *I-V* characteristics for various salt solutions (color-coded) using the reference electrodes; $\Delta C$ = 10. The bottom inset shows similar measurements but with standard Ag/AgCl electrodes. Top inset: drift-diffusion experimental setup.

When using reference electrodes, the measured zero current potential gives $E_m$ directly. However, we often found it beneficial to use the standard Ag/AgCl electrodes (without immersion into a saturated KCl/NaCl solution), instead of the reference electrodes. The use of such electrodes is known to result in an additional voltage drop of ~55 mV, which was close to the expected values $E_{redox}$ given by[12]

$$E_{redox} = \frac{k_B T}{e} ln\left(\frac{\gamma_H}{\gamma_L}\Delta C\right) \quad (S1)$$

where $\gamma_H$ and $\gamma_L$ are the mean activity coefficients of the ions in the high and low concentration solutions, respectively. These coefficients are known to depend slightly on chosen salts and their concentrations[12,15]. In the case of the standard Ag/AgCl electrodes, $E_m$ could then be obtained from the measured zero-current voltage $E_{total}$ by subtracting the literature values of $E_{redox}$ as

$$E_m = E_{total} - E_{redox} \quad (S2)$$

The zero-current potentials $E_m$ measured using either reference or Ag/AgCl electrodes agreed within our experimental accuracy of ± 1 mV. The obtained $E_m$ for each studied salt is given in Supplementary Table 2. From those $E_m$, we could evaluate the mobility ratio $\mu^+/\mu^-$ for cations and anions in the bulk using the Henderson formula (Eq. 1), as discussed in the main text. The obtained ratios are given in Supplementary Table 2 and compared with the values reported in the literature[12].

For Å-channel devices, we performed *I-V* and drift-diffusion measurements of respective salts in similar way as described for microhole, and obtained the mobility of each ion. The average of the obtained values of the individual ion mobilities (further details on mobility calculation in supplementary section 4) from each device is presented, whereas the sample-to-sample variations are reflected in the error bars obtained from the standard deviation of all the devices.

**Supplementary Table 2|** Membrane ($E_m$) and total ($E_{total}$) potentials for salt solutions as measured using microhole devices by the reference and standard Ag/AgCl electrodes, respectively.

| Salts | Experimental $E_m$ (mV) | Experimental $E_{total}$ (mV) | Expected[12] $E_{redox}$ in mV | Experimental Microhole $\frac{\mu^+}{\mu^-}$ | Bulk $\frac{\mu^+}{\mu^-}$ (ref 12) |
|---|---|---|---|---|---|
| KCl | 1.5 | 56 | 55 | 1.05 | 0.96 |
| CsCl | 1.6 | 56 | 54 | 1.05 | 1.01 |
| KF | 9 | 64 | 55 | 1.36 | 1.32 |
| KI | 0.9 | 56 | 55 | 1.03 | 0.95 |
| KClO$_4$ | 5.6 | 62 | 55 | 1.21 | 1.09 |
| NaCl | -9 | 46 | 55 | 0.73 | 0.65 |
| LiCl | -16 | 40 | 56 | 0.57 | 0.51 |

**Supplementary Section 4. Individual ion mobility calculation in experiments**

The individual ion mobilities are obtained from the $E_m$ through the combined use of Henderson's equation (Eq. 1 in the main text), and the relation between conductivity and mobility, $\sigma = eN_A C (\mu^+ + \mu^-)$. This data analysis is done assuming the concentration of ions inside nanochannels and bulk are similar. We acknowledge that the concentration gradient can appear for several reasons including access resistance and a large contribution of the surface conductance[16-18]. We verify that the access resistance is negligible from the resistance versus channel length $L$ plots (main Fig. 1c inset) where the linear fits extrapolate to zero for three different ionic solutions (KCl, KF and KClO$_4$). However, this assumption of little to no concentration gradients (and the use of the Henderson equation) may not be necessarily applicable for even longer (e.g., 10 times) channels than those employed in our study. As per our theoretical simulations, let us note that for the channels with $h$ = 6.8 Å, the free energy of an ion along the channel length direction differs only by few k$_B$T when compared to the bulk for monovalent ions[21]. For bivalent ions, or for the channels much thinner than the $D_H$, the energy barriers could be much larger [21], and the concentration difference between bulk and the channel is also important. For this reason, we deliberately use monovalent ions in our study, with the channel height comparable to that of the ions' $D_H$. In line with our study, the modification of the speed of ions rather than their concentrations in nanochannels was observed by Ma et al. in atomistic simulations for nanochannels[20].

**Supplementary Section 5. Molecular dynamics simulations**

To understand the observed influence of ion-core and hydration diameters on ions' transport through Å-channels, we performed equilibrium molecular dynamics (MD) simulations. Our setup consisted of two parallel graphene sheets that created a 2D capillary with the effective height $h$ = 6.8 Å, as in our experiments. The channel length was chosen to be 46.2 Å. The carbon atoms were fixed in their positions. The space between the graphene sheets was filled with virtual aqueous solutions of KCl, CsCl, NaCl, LiCl, KClO$_4$, KF or KI in 1M concentration. The SPC/E model[22] was used to describe interactions between water molecules whereas interaction between ions and water included both Coulombic and van der Waals contributions. Van der Waals interactions including those between water and confining walls were described by the Lennard-Jones (LJ) potential. The used parameters for LJ potential and SPC/E model are listed[23,24] in Supplementary Table 3. The Lorentz–Berthelot combination rule[25,26] was used for the interaction between different atoms, i.e. $s = (s_i + s_j)/2$ and $\varepsilon = \sqrt{\varepsilon_i \varepsilon_j}$ where $i$ and $j$ are two different elements. The MD simulations were performed using LAMMPS[27] with periodic boundary conditions applied in all three directions. The short-range interactions were truncated with a cutoff at 12 Å whereas long-range (Coulomb) forces were computed by utilizing the particle–particle particle–mesh algorithm. The system was equilibrated using isothermal−isobaric ensembles for 10 ns with a timestep of 1 fs. Temperature $T$ was fixed at 300 K using the Nosé−Hoover thermostat. We used canonical ensembles to evaluate observables in these simulations[28,29].

**Supplementary Table 3|** The Lennard-Jones parameters and charges for each element in the molecular dynamics simulations.

| element | s (Å) | ε (kcal/mol) | Q (e) |
| --- | --- | --- | --- |
| O (in water) | 3.166 | 0.156 | -0.8476 |
| H (in water) | 0 | 0 | +0.4238 |
| F | 3.117 | 0.1794 | -1 |
| Cl | 4.401 | 0.1 | -1 |
| I | 5.167 | 0.1 | -1 |
| Cl in ClO$_4$ | 4.9 | 0.04 | +1.309 |
| O in ClO$_4$ | 3.1 | 0.075 | -0.577 |
| K | 3.331 | 0.1 | +1 |
| Cs | 3.883 | 0.1 | +1 |
| Na | 2.583 | 0.1 | +1 |
| Li | 1.505 | 0.164 | +1 |

In addition to Fig. 3 of the main text, examples of the calculated presence probabilities are shown in Supplementary Fig. 5 for F$^-$ and K$^+$ ions that tend to reside in the channel's center. One can see that their probabilities are well described by the Gaussian distribution. Furthermore, the probability of ions position data for the ions Na$^+$ and I$^-$ were not shown in Fig. 3 of the main text for the sake of the figure's clarity. Accordingly, they are presented in Supplementary Fig. 5c.

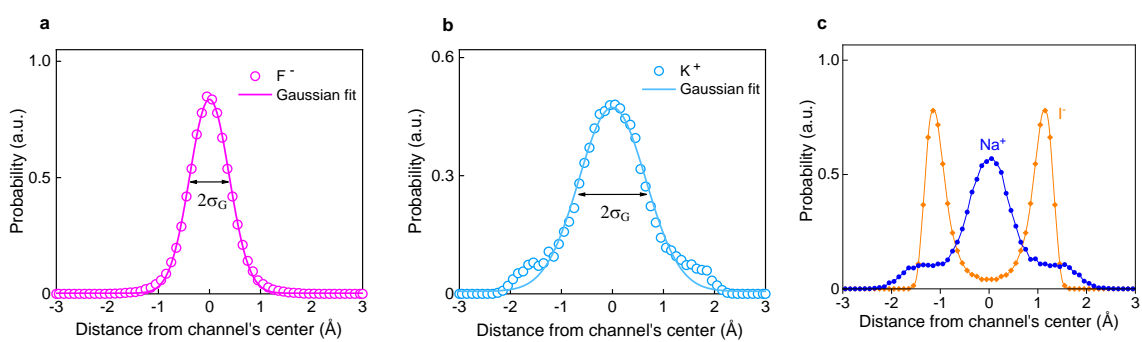

**Supplementary Fig. 5|** Positions of different ions in bilayer water inside our Å-channels. (**a** and **b**) Presence probabilities of fluoride and potassium ions, respectively. Solid curves: best Gaussian fits

that yield the central (zero) position and the variances of ~0.38 and 0.65 Å for $F^-$ and $K^+$, respectively. (**c**) Probabilities for $Na^+$ and $I^-$ which were not presented in Fig. 3b of the main manuscript, for clarity.

To provide an independent proof for different stable positions of different ions inside the 2D capillaries, we performed complementary MD simulations and calculated the potential of mean force (PMF) that is a measure of the free energy along the *z* direction[30]. The results are shown in Supplementary Fig. 6 for $F^-$ and $Cl^-$. It is clear that the free energy for $F^-$ has a minimum in the channel's center whereas $Cl^-$ has its minimum off-center, inside one of the water layers. The found free energy minima are close to the calculated maxima in the presence probability (shaded green regions in Supplementary Fig. 6).

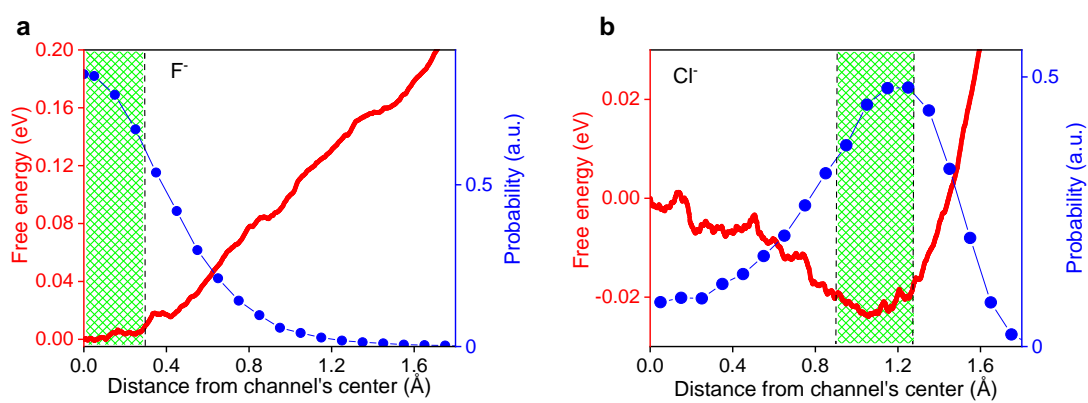

**Supplementary Fig. 6| Free energy calculations.** The free energy for fluoride (**a**) and chloride (**b**) ions inside the Å-channels (red curves), and their presence probability (blue). The shaded areas indicate the spatial range where the minima of the free energy coincide with the maxima of the presence probability.

To understand the effect of capillary walls' materials, the simulations were also carried out for an additional wall material i.e., hBN channels of the same 6.8 Å height by replacing the graphene monolayers with hBN ones. Simulation parameters for hBN were taken from ref.[31]. We did not find any notable differences in ion distributions inside the channels made from two different 2D materials, as shown in Supplementary Fig. 7. This indicates that the key role in ions' positions and, hence ion selectivity, is played by the structure of bilayer water rather than ion interactions with the surfaces. The distribution of water molecules within the channels is shown in Supplementary Fig. 8.

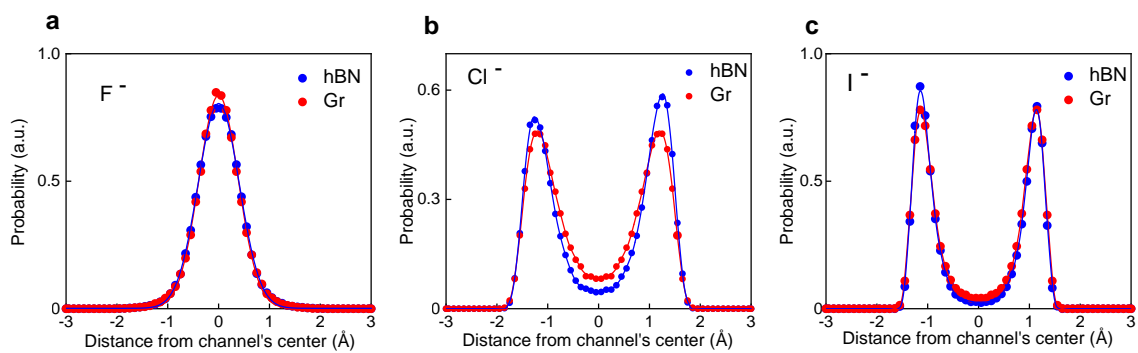

**Supplementary Fig. 7| Ions' positions in channels with different wall materials.** Presence probabilities of (**a**) fluoride, (**b**) chloride, and (**c**) iodide ions inside graphene and hBN channels with *h* = 6.8 Å. MD simulations were done for 1M solutions.

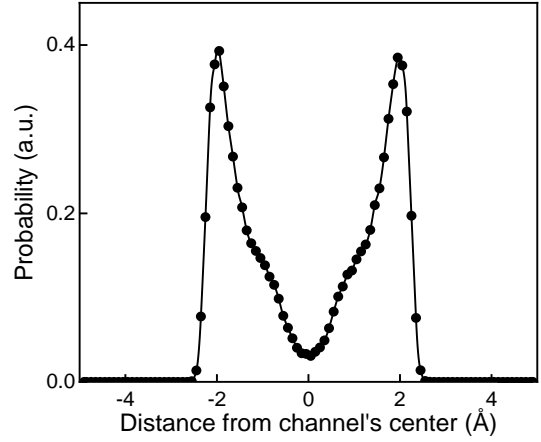

**Supplementary Fig. 8| Distribution of water molecules inside Å-channels.** Presence probability of water confined in Å-channels with graphene walls, from MD simulations. The two peaks reaching maxima at ± 2 Å correspond to the two layers of water within the graphene-confined channel. These results were similar for the channels made from hBN.

**Supplementary Section 6. Mobility and diffusion coefficient from MD simulations**

The diffusion coefficient $D$ was calculated using the mean-squared displacement method as

$$D = \frac{1}{2\Phi t}<(r(t) - r(0))^2> \qquad (S3)$$

where $r(t)$ is the position of an ion at the time $t$ and $\Phi$ is the dimensionality of its diffusive trajectory. We used $\Phi$ = 2 and 3 for our Å-channels and the bulk, respectively. The brackets in Eq. S3 refer to time averaging or an average over many trajectories, which can be found by performing MD simulations in both cases[32]. Usually, we calculated a time average using a single long (20 ns) ion trajectory. The MD simulations were done for the bulk density of water and 1 M salt concentrations. From the diffusion coefficients found using Eq. S3, we calculated ion mobilities $\mu$ employing the Einstein relation

$$\mu = \frac{e}{k_B T} D \qquad (S4)$$

where $e$ is the monovalent-ion charge, $k_B$ is the Boltzmann constant and $T$ = 300 K. We repeated the simulation 10 times using different seed points for each simulation, and calculated the error from the standard deviation of the data.

MD simulations are known to suffer from systematic errors in describing absolute values of transport coefficients. Even for ion mobility in bulk, discrepancy of up to 30% was observed between theory and experiments [23]. There could be several reasons for this difference as the ion's mobility is a complicated function of ion's charge and size, modulated by the structure and dynamics of the hydration shells and the solvent. However, MD simulations are good at capturing the general trends when a system's parameters change [23]. Accordingly, to minimize an effect of systematic errors in our analysis, we used the relative mobilities $\mu_{Å\text{-channel}}/\mu_{bulk}$ when comparing our experimental results with the simulations (Fig. 3d of the main text). The absolute values of the mobilities along with error values obtained in the experiments and MD simulations for each studied salt is given in Supplementary Table 4. There are similar trends of mobility for different ions, as seen in comparison of the mobility from MD simulations and experimental data in Supplementary Fig 9.

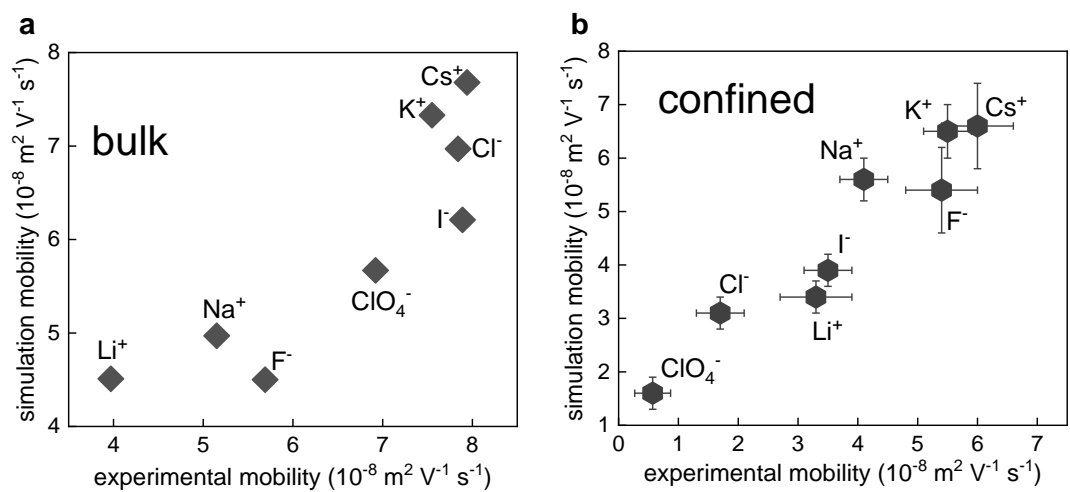

**Supplementary Fig. 9| MD simulations against experimental mobility data.** Individual ion mobility values depicts similar trends in experiments and simulations for (**a**) bulk ion mobility, and (**b**) confined ion mobility. For the bulk values in **a**, the symbol size represents the error. In **b,** the horizontal error bars for experimental data are from standard deviation (±SD) of the ion mobility from 3 devices, and for simulations data, the vertical error bars were obtained from ±SD of over ten simulations; each has different initial seed of velocities drawn from the Maxwell-Boltzmann distribution.

**Supplementary Table 4|** The mobility of individual ions in confined and bulk systems obtained from experiments and MD simulations. For the bulk values, the error is less than 0.1 (x $10^{-8}$ m$^2$/V/s).

| Ion | Experiment | MD simulation | Experiment | MD simulation |
|---|---|---|---|---|
| | bulk mobility ($10^{-8}$ m$^2$/V/s) | bulk mobility ($10^{-8}$ m$^2$/V/s) | confined mobility ($10^{-8}$ m$^2$/V/s) | confined mobility ($10^{-8}$ m$^2$/V/s) |
| F$^-$ | 5.69 | 4.5 | 5.4 ± 0.6 | 5.4 ± 0.8 |
| Cl$^-$ | 7.84 | 6.97 | 1.7 ± 0.4 | 3.1 ± 0.3 |
| I$^-$ | 7.89 | 6.21 | 3.5 ± 0.4 | 3.9 ± 0.3 |
| ClO$_4^-$ | 6.92 | 5.67 | 0.6 ± 0.3 | 1.6 ± 0.3 |
| Li$^+$ | 3.97 | 4.51 | 3.3 ± 0.6 | 3.4 ± 0.3 |
| Na$^+$ | 5.15 | 4.97 | 4.1 ± 0.4 | 5.6 ± 0.4 |
| K$^+$ | 7.55 | 7.33 | 5.5 ± 0.4 | 6.5 ± 0.5 |
| Cs$^+$ | 7.94 | 7.68 | 6.0 ± 0.6 | 6.6 ± 0.8 |

**Supplementary Section 7. Effects of walls and counterions**
To further elucidate the effect of ions' position inside Å-channels on their permeation, Supplementary Fig. 10b shows the measured mobilities of the studied ions as a function of (a) their DI and (b) the probability to be found near channel walls (that is, > 1.0 Å away from the center). Most of the ions (except for Li$^+$) exhibit a clear correlation between $\mu$ and the near-wall presence. Ions with large $D_I$ such as Cl$^-$, I$^-$ and ClO$_4^-$, which stay near the walls most of the time, were found to have strongly suppressed mobility with respect to the bulk values. On the other hand, F$^-$ tending to stay in the center shows little mobility suppression.

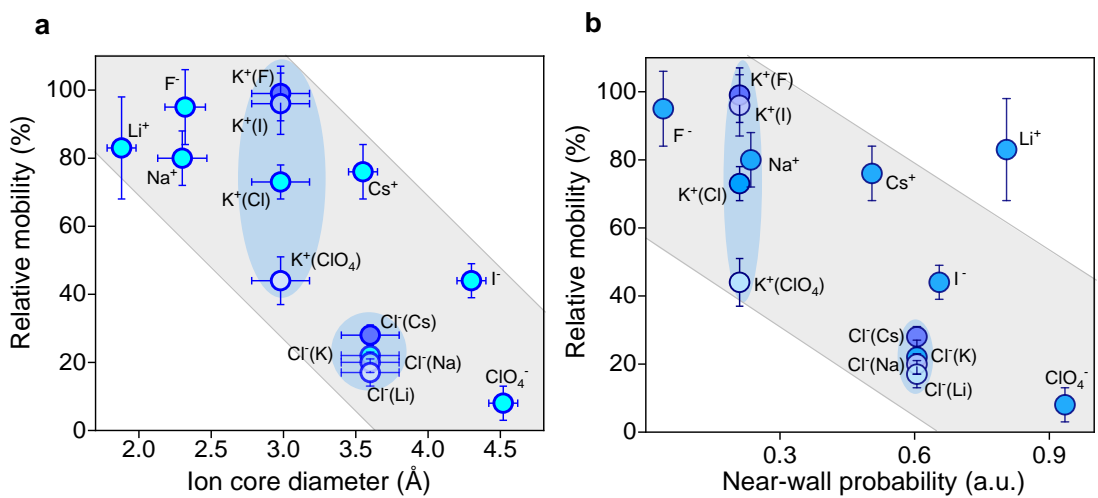

**Supplementary Fig. 10| Influence of ion core size and near-wall positions of ions on their mobility.** Experimental measured relative mobilities of Å-confined ions (with respect to the bulk values) as a function of (**a**) ionic diameter, $D_I$ and (**b**) the probability of finding them near channel walls. The mobilities are given in % relative to the bulk values. In (a) and (b) K$^+$ and Cl$^-$ mobilities were measured for four different counter-ions, as indicated in the parentheses. The shaded areas are guides to the eye. Vertical error bars in **a** and **b**, experimental data from the average ± SD of the ion mobility from 3 devices. The horizontal error bars in (a) indicate the spread in the $D_I$ values from the literature[7,8].

In bulk solutions, ions and their counter-ions diffuse independently and, for example, cations' mobility is not affected by mobility of anions. On the other hand, certain influence of counterions on mobility of the same ion has been reported for biological channels[33]. Such interaction effects remain poorly studied for the case of artificial channels[34]. Comparing the potassium salts with different anions, K$^+$ ion mobility was dependent on the counter-ion (Fig. 2d in main and Supplementary Fig 10).

**Supplementary Section 8. Angular orientations of water molecules around ions**

To study the angular orientations of water molecules around ions, we first define the included angle θ between the line segment which connects an ion and the oxygen atom in the first hydration shell, and z axis (any direction in the bulk solution or the direction perpendicular to the graphene walls). Schematic diagrams are shown in Supplementary Fig 11. Then we calculated the probability distribution of θ in the first hydration shell, for both Cl$^-$ and K$^+$, in bulk solution (Supplementary Fig. 12a and b) and Å-channel (Supplementary Fig. 12c and d), respectively.

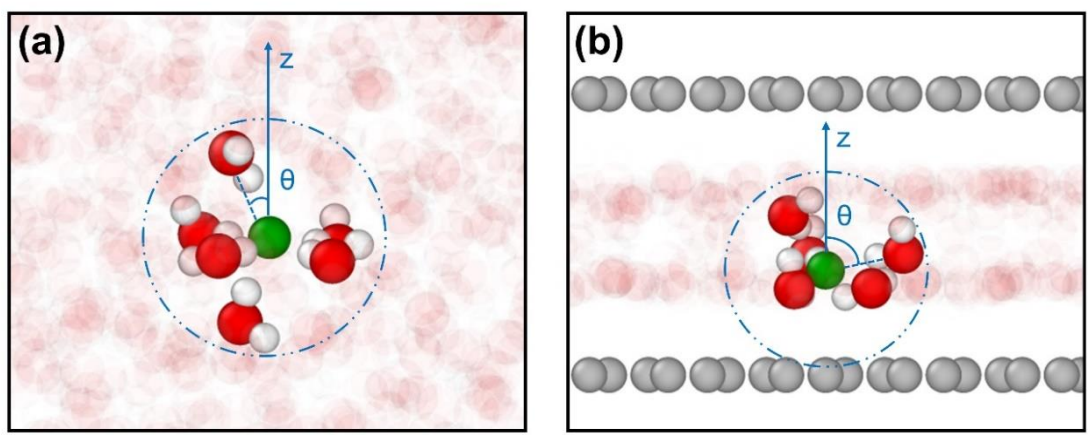

**Supplementary Fig. 11| Angular orientation of water molecules around ions.** Schematics of hydrated ions in (**a**) the bulk solution and (**b**) the Å-channel. The $Cl^-$ ions are represented by green balls. H and O of $H_2O$ are represented by white and red balls respectively. Blue dashed circles are the first hydration shells. Carbon atoms of graphene are represented by grey balls. Here θ is water orientation angle and *z*-axis represents the direction relative to which the angle is defined.

For uniformly distributed water molecules around ions (i.e., in the bulk solution), the probability density of θ is (½)sin θ. Our MD results agree well with the theoretical fitting, as shown in Supplementary Fig. 12a and b. However, in the Å-channel, the hydration shells have to adapt to the confined bilayer water and cannot maintain spherical structures as they do in bulk solutions. We have shown that $K^+$ ions are more likely to locate at the centre of the channel, whereas $Cl^-$ ions are located near the walls of the channel[35]. Consequently, the distribution of water molecules around ions is not uniform. Our MD results reveal that the probability distribution of θ in the Å-channel manifests a significant difference compared to that in the bulk solution, as shown in Supplementary Fig.12c and d. The peaks can be interpreted by the uneven distribution of oxygen atoms of the confined bilayer water.

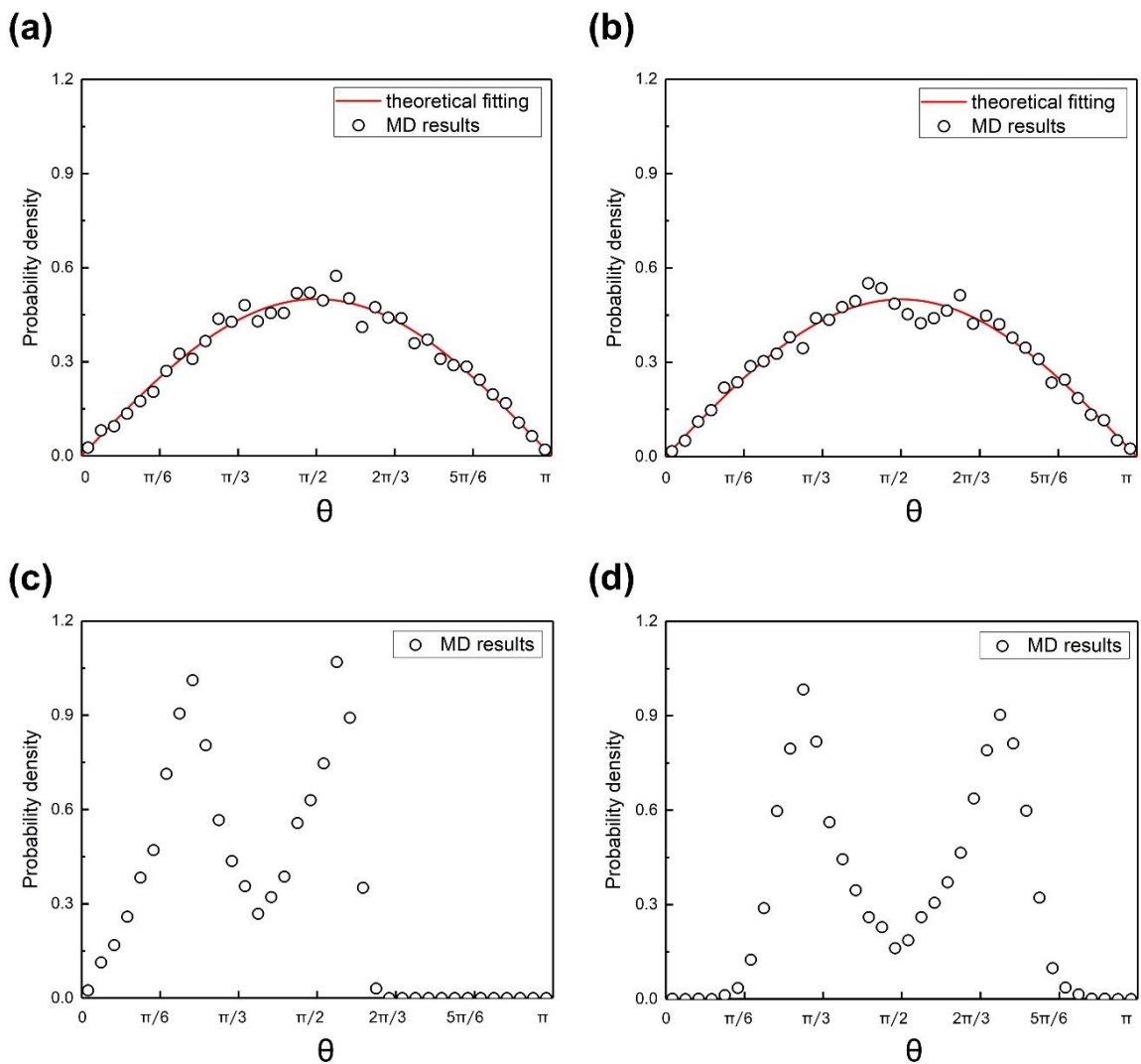

**Supplementary Fig. 12| Probability of angular orientation.** Probability distribution of θ, i.e., angular orientations of water molecules around ions in bulk solutions of (**a**) Cl$^-$, and (**b**) K$^+$, and in confined Å-channel for (**c**) Cl$^-$, and (**d**) with K$^+$. Black circles are probability densities of different values of θ from MD simulations. Red lines give the theoretical fitting using $\frac{1}{2}\sin\theta$.

**Supplementary Section 9.** *Ab initio* **simulations for determining hydration shell and energy**

We performed quantum simulations to study the effect of ions' Å-confinement on the arrangement of the immediate water molecules for anions and cations in bulk and confined systems. Depending on the channel size, for channels with $h \geq D_H$, ions only lose a few of their inner hydration water molecules when they enter the channel. When the size of the channel is smaller than the radius of the first hydration layer, further dehydration occurs. For the effective height of channel h = 6.8 Å, the hydration layers stay partially intact except for the portion of the hydration surface cut off by the channel walls. In other words, the portion of the hydration layer close to the channel's wall changes significantly. One can consider a set of surfaces placed at each hydration layer, i, at radius $R_i$. Each surface represents the area where the water dipoles fluctuate, giving the time-averaged dipole layers[36]. We calculated hydration energy ($E_H$) per water molecule of different ions inside and outside of the channel which are presented in Supplementary Table 5.

We considered that six water molecules are around the ions of alkali metal and halogen atoms, which is considered to be the first hydration shell[37-39]. The optimized structures of the hydrated ions with six water molecules are shown in Supplementary Fig. 13 and 14, indicating that the water molecules are oriented from the O and H side around the cations and anions, respectively. It was reported that in the first hydration shell, the dipole moment of water molecules decreases as the number of water molecules increases in relation to the suppression of the ion electric field[36,40]. The latter leads to a different orientation of the water molecules in the next hydration shell of ions compared to that of the first hydration shell. The NBO (natural bond orbitals) data provided in Supplementary Fig. 13 and 14 indicate that depending on the ion type and its distance from the wall, charge transfer that takes place from ions to water molecules for anions and for cations can be more complex. In an ideal situation Q=-1e for anions and Q=+1e for cations. We need to consider two different levels of calculation (quantum and classical rigid models) in different frameworks, i.e., the electronic nature of water-ion interaction should be studied using quantum mechanical calculations, while the ordinary thermodynamic properties can be obtained using rigid water models.

The $E_H$ is estimated using the following equation:

$$E_H = E_T - (E_w + E_i + E_g) \qquad (S5)$$

where $E_T$, $E_w$, $E_i$, $E_g$, is the total energy of the system, water molecules, and ion and graphene sheets, respectively. Note that for bulk, $E_g$=0. Our ab-initio result is comparable to the results obtained from the model proposed in Ref. 36. For instance, the solvation energy for Cl$^-$ anion per water molecule was reported to be -1.7/6.0 = -0.28[36] for the first layer which is two times smaller than our obtained number, i.e., -0.68 eV.

The charge distributions on cations and anions hydrated by six water molecules are visualized in electrostatic potential (ESP) energy maps shown in Supplementary Fig. 13 and 14 (right hand side, RHS). The positive regions (blue) of the ESP maps are related to the distribution of positive charge and (nucleophilic reactivity) and the negative regions (red) to the distribution of negative charge (electrophilic reactivity). Accordingly, the charge distribution in the nearest molecules of the hydrated ions decreases as the size of the ions increases.

The polarized water molecules in the hydration shells can be affected by the ion charge. It is known that OH$^-$ groups interact with hydrophobic walls much stronger[41,42] than H$^+$, which can result in a stronger influence of the walls on the diffusion of ions depending on whether they point OH$^{\delta-}$ or H$^{\delta+}$ toward the walls. Since classical molecular dynamics simulations cannot describe such subtle details for ion interactions with channel walls, we employed density functional theory (DFT) calculations to explain our observations quantitatively. For instance, the water molecules around the Cl anion in the bulk orient themselves in such a manner that single hydrogen from each molecule points toward the ion (see RHS-bottom Supplementary Fig.13b). The innermost layer is very tightly bound (Supplementary Fig. 13), and subsequent layers are spaced at longer distances. If the ion is placed in the Å-channel, the innermost hydration layer is affected, see LHS panels in Supplementary Fig. 13 and 14. For thicker channels, the hydration layer structure is identical to that in bulk water. As we see from Supplementary Fig. 13 and 14 the coordination number decreases when the ion is inside the narrow channels of 6.8 Å (see Supplementary Table 5). In order to find the coordination number, we first optimized 6 water molecules around each ion in the bulk and found the distance between each water molecule and the central ion as shown by dashed circle in the Supplementary Fig. 13 and 14. Then, using that distance, we counted number of water molecules around ion in confined case, which gave us the coordination number for confined ions. Thus, coordination number is interlinked with this distance we fixed between water and the ion in the bulk. We also reported the charges over ions and hydration energies for various ions in Supplementary Table 5.

**Supplementary Table 5|** DFT simulations: The values below in the table are for 1) Q, the charge over ions; 2) $E_H$, hydration energy; 3) coordination number; 4) *d*, the off-centre distance from the centre of the ion to the channel centre. In the confined channel, the coordination number is measured when the shortest distance between ions and water molecules are set to be the distance they have in bulk case. The off-centre distance d is schematically depicted in Supplementary Fig. 15.

| | $\xrightarrow{M}$ | $Li^+$ | $Na^+$ | $K^+$ | $Cs^+$ | $F^-$ | $Cl^-$ | $I^-$ | $ClO_4^-$ |
|---|---|---|---|---|---|---|---|---|---|
| Q(e) | Bulk | +0.64 | +0.72 | +0.93 | +0.91 | -0.73 | -0.87 | -0.91 | -0.86 |
| | Channel | +0.57 | +0.74 | +0.84 | +0.88 | -0.69 | -0.76 | -0.78 | -0.87 |
| $E_H$(eV) | Bulk | -1.01 | -0.77 | -0.47 | -0.09 | -0.91 | -0.68 | -0.55 | -0.60 |
| | Channel | -0.37 | -0.33 | -0.27 | -0.08 | -0.31 | -0.27 | -0.26 | -0.18 |
| coordination number | Bulk | 6 | 6 | 6 | 6 | 6 | 6 | 6 | 6 |
| | Channel | 4 | 4 | 4 | 3 | 4 | 4 | 4 | 3 |
| d(Å) | | -0.68 | +0.16 | -0.40 | -1.53 | +0.36 | -1.66 | -0.77 | -1.99 |

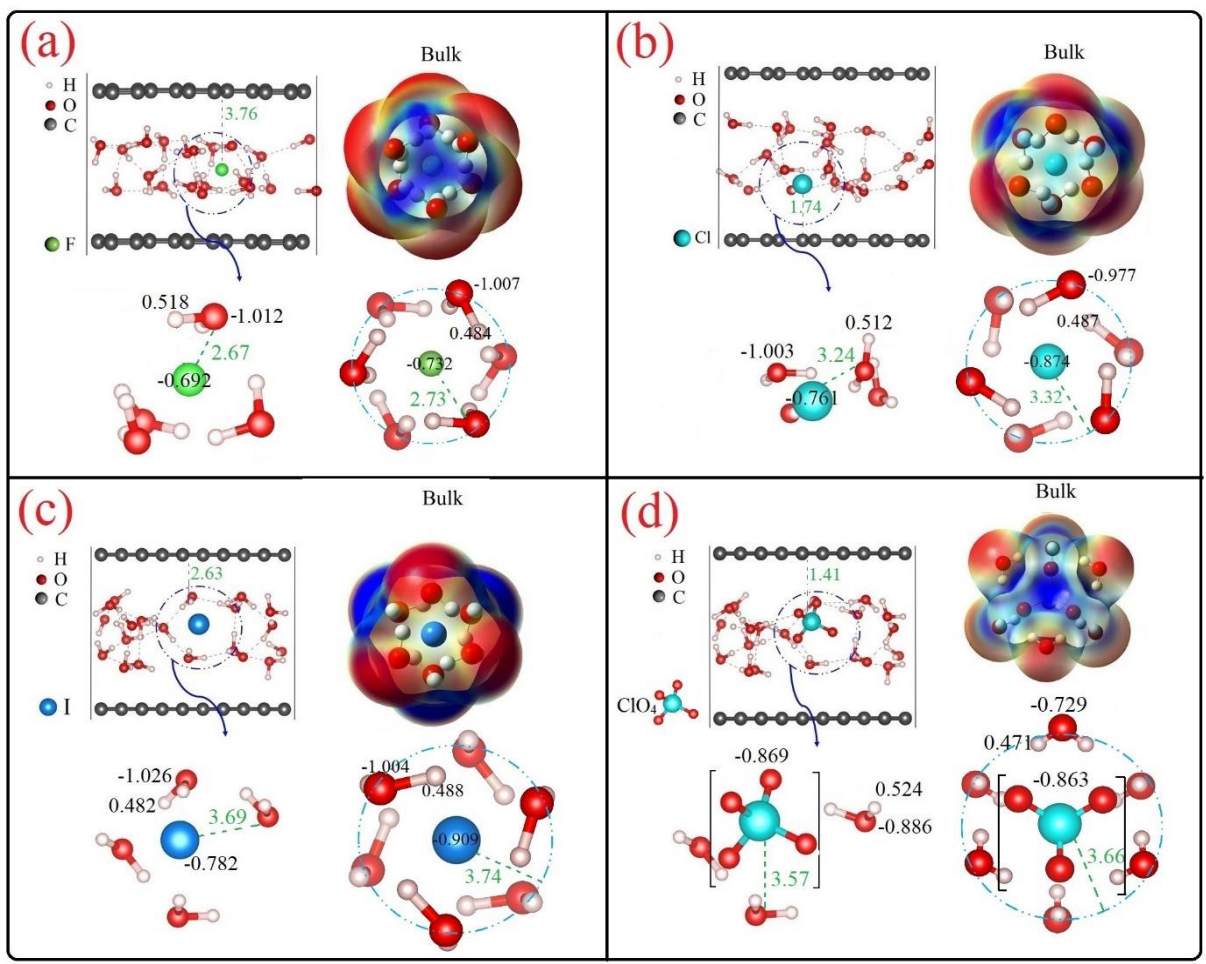

**Supplementary Fig. 13| Equilibrium geometries and electrostatic potential (ESP) map of the hydrated anions**. The optimized structures and the natural bond orbitals (NBO) charges (black numbers) and ion distance to the water molecule (green numbers) of hydrated anions (**a**) F$^-$, (**b**) Cl$^-$, (**c**) I$^-$, and (**d**) ClO$_4^-$ in bulk with six water molecules (right hand side, RHS-bottom of each panel) and the hydrated anions between graphene layers (left hand side, LHS-bottom is zoomed in structure of each panel). In each panel, RHS-up is the electrostatic potential energy maps of the hydrated anions. Electrostatic potentials are mapped on the surface of the electron density of 0.002 units.

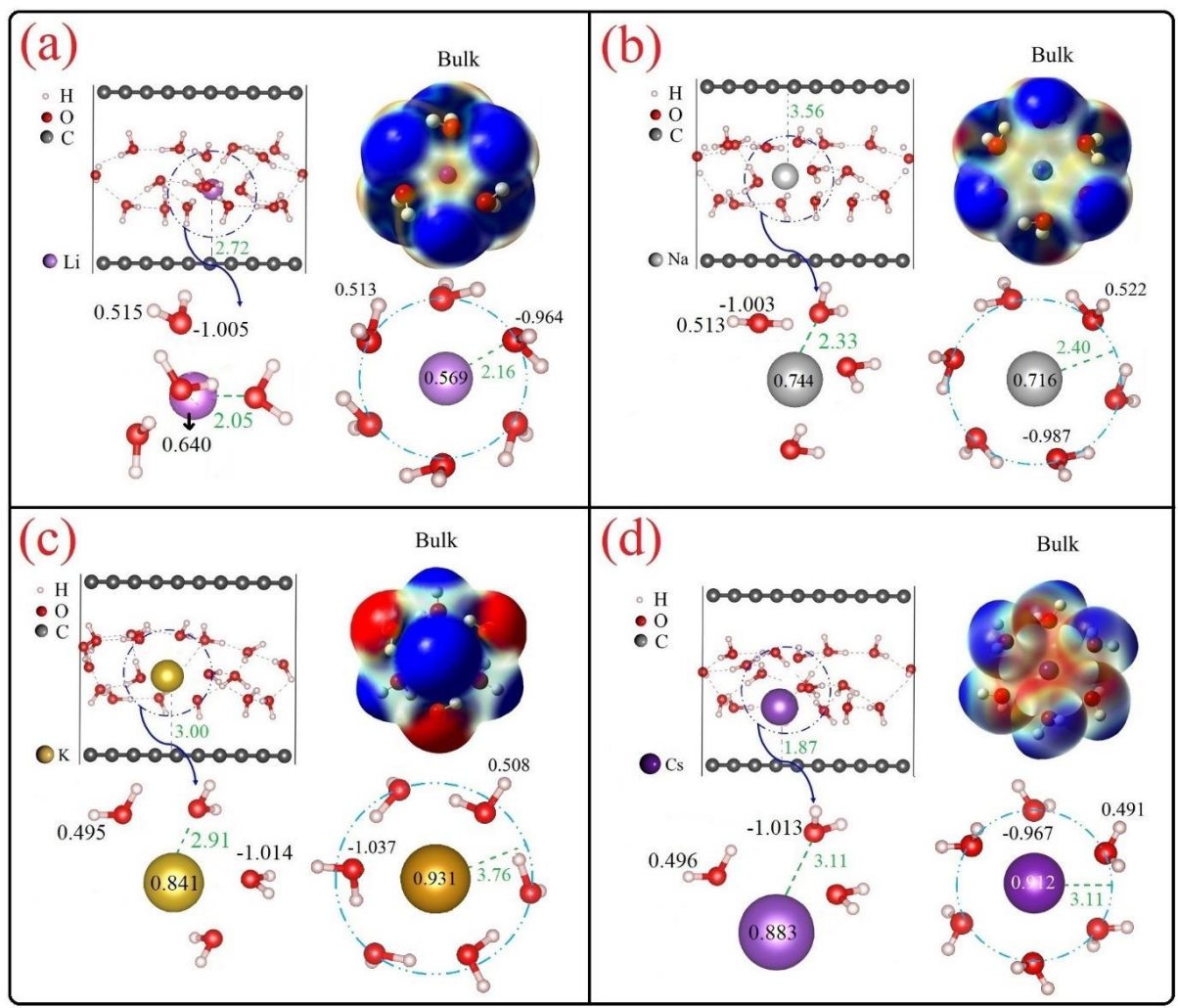

**Supplementary Fig. 14| Equilibrium geometries and ESP map of the hydrated cations**. The optimized structures and the NBO charges (black numbers) and ion distance to the water molecule (green numbers) of hydrated cations (**a**) $Li^+$, (**b**) $Na^+$, (**c**) $K^+$, and (**d**) $Cs^+$ in bulk with six water molecules (RHS-bottom of each panel) and the hydrated cations between graphene layers (LHS-bottom is zoomed in the structure of each panel). In each panel, RHS-up is the electrostatic potential energy maps of the hydrated cations. Electrostatic potentials are mapped on the surface of the electron density of 0.002 units.

**Method details - Equilibrium geometries of the hydrated ions between two graphene layers**
We performed DFT calculations using the DMol$^3$ module of the Materials Studio software [43]. The DFT calculations were performed based on the generalized gradient approximation (GGA) with Perdew-Burke-Ernzerhof (PBE) functional and the double numerical plus polarization (DNP) basis set[44,45]. Long-range vdW interactions were taken into account using Grimme's van der Waals correction scheme [46]. The supercell cell was two graphene layers with a size of 12.37 Å ×12.86 Å that were separated by an effective height ($h$) of 6.8 Å. The interlayer space of two graphene sheets was filled with 20 water molecules and an ion. The graphene layers were fixed in their initial positions in all DFT calculations and to avoid the interaction of two adjacent layers, a vacuum layer of 20 Å was inserted along the *z-axis*. The global orbital cut-off and the self-consistent field (SCF) convergence threshold were set as 6.0 Å and $10^{-6}$ a.u., respectively.

The optimized structures of the hydrated ions between the graphene layers are shown in Supplementary Fig. 13 and 14. Also, the off-centre distance of cations ($M^+$, M=Li, Na, K, Cs) and anions ($N^-$, N= F, Cl, I, $ClO_4$) from the centre of the channel (Supplementary Fig. 15) are tabulated in Supplementary Table 5. In good agreement with MD results, the $Na^+$, $K^+$, and $F^-$ ions are located near the center of the channel. The $I^-$, $Cs^+$, $ClO_4^-$, and $Cl^-$ ions with large ionic radii, prefer to be closer to the walls. In case of the $Li^+$ with an ionic radius of 0.60 Å, its distance from the center is 0.68 Å and it is close to the wall.

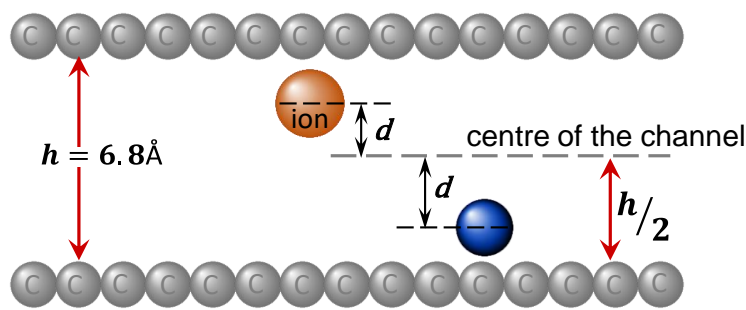

**Supplementary Fig. 15| Ions' off-centre distance.** Schematic depicting the off-centre distance of an ion, which is measured from the centre of the channel, to the centre of the ion. The height of the channel is *h*, and *h*/2 is the distance to centre of the channel. Blue and orange colored balls represent ions with two typical locations, i.e., close to the wall or close to the centre respectively. The distance of the ion '*d*' values from the centre of the channel to the centre of a particular ion is the off-centre distance.

**Supplementary references**